\documentclass[twocolumn]{aastex631}
\usepackage{stfloats}
\usepackage{float}
\usepackage{graphicx}
\usepackage{natbib}
\usepackage{hyperref}
\usepackage{amsmath}
\usepackage{autobreak}
\usepackage{multirow}
\usepackage{makecell}
\usepackage{color}
\usepackage{bm}
\usepackage{threeparttable}
\usepackage[figuresright]{rotating}
\usepackage[mathscr]{euscript}
\usepackage[mathlines]{lineno}
\usepackage[utf8]{inputenc}
\usepackage[T1]{fontenc}
\usepackage{amsmath}
\usepackage{makecell}

\begin{document}

\title{The disappearance of the blue and luminous progenitor of the Type IIn SN~2010jl}

\correspondingauthor{Ning-Chen Sun}
\email{sunnc@ucas.ac.cn}

\author[0000-0002-3651-0681]{Zexi Niu}
\affiliation{School of Astronomy and Space Science, University of Chinese Academy of Sciences, Beijing 100049, People's Republic of China}
\affiliation{National Astronomical Observatories, Chinese Academy of Sciences, Beijing 100101, China}

\author[0000-0002-4731-9698]{Ning-Chen Sun}
\affiliation{School of Astronomy and Space Science, University of Chinese Academy of Sciences, Beijing 100049, People's Republic of China}
\affiliation{National Astronomical Observatories, Chinese Academy of Sciences, Beijing 100101, China}
\affiliation{Institute for Frontiers in Astronomy and Astrophysics, Beijing Normal University, Beijing, 102206, People's Republic of China}

\author{Jifeng Liu}
\affiliation{National Astronomical Observatories, Chinese Academy of Sciences, Beijing 100101, China}
\affiliation{School of Astronomy and Space Science, University of Chinese Academy of Sciences, Beijing 100049, People's Republic of China}
\affiliation{Institute for Frontiers in Astronomy and Astrophysics, Beijing Normal University, Beijing, 102206, People's Republic of China}
\affiliation{New Cornerstone Science Laboratory, National Astronomical Observatories, Chinese Academy of Sciences, Beijing 100012, People's Republic of China}

\begin{abstract}

Type IIn supernovae (SNe) exhibit narrow hydrogen lines that arise from the strong interaction between ejecta and circumstellar material. It remains poorly understood, however, what progenitor stars give rise to these explosions.  
In this work, we perform a detailed analysis of the progenitor and environment of the nearby Type IIn SN~2010jl. With newer images taken by the Hubble Space Telescope, we confirm that the previously reported progenitor candidate is a blend of the progenitor itself and a field star cluster in its close vicinity. SN~2010jl has now become much fainter than the progenitor. 
The progenitor is very blue and luminous with an effective temperature of log $T_{\rm eff}/{\rm K}$=4.26$^{+0.11}_{-0.09}$ and a luminosity of log $L/L_{\odot}$ =6.52$^{+0.20}_{-0.16}$. It is located in a very young star-forming region, but its luminosity is much higher than that expected from the environmental stellar populations. We suggest that the progenitor was in outburst when observed. Its nature and evolutionary history remain to be investigated. 

\end{abstract}

\section{Introduction} \label{sec:intro}

Type IIn supernovae (hereafter SNe~IIn) is an interesting and important subclass of core-collapse SNe. 
They are characterized by strong and narrow (a few hundreds to about 2000 km s$^{-1}$) emission lines, sometimes with broad bases, in the spectra \citep{1990Schlegel}.
These features indicate dense circumstellar material (CSM) surrounding the SN,
which requires enhanced mass loss (10$^{-3}$--10$^{0}$ $M_{\odot}$/years ) of the progenitor stars within a short timescale prior to the deaths. Therefore, SNe~IIn are an effective tool to investigate the mass loss of massive stars at their final evolutionary stage.

The progenitor of SNe~IIn is a long-standing question.
The connection between SNe~IIn and luminous blue variables (LBVs) is widely speculated and supported by the direct progenitor detections, such as SN~2005gl, SN~2009ip, SN~2010bt, and SN~2015bh \citep{2016Elias-Rosa, 2018EliasRosa, 2009GalYam, 2010Smith}. The detected progenitors have prominent pre-explosion variabilities and extremely high luminosities that even exceed the Humphrey–Davidson instability limit \citep{1994Humphreys}, both features reminiscent of known Galactic LBVs such as $\eta$~Carinae ($\eta$~Car). 
However, LBVs are not expected to explode in the traditional view, since it was believed to be a transitional stage of very massive single stars (30--100 $M_\odot$) toward the Wolf-Rayet phase (\citealp{1994Humphreys,1996GarciaSegura}; see review in \citealp{2017Smith}).
Meanwhile, other hypotheses have also been proposed to explain SNe~IIn, including pulsational pair-instability SNe of very massive stars (SN~2006gy, \citealp{2007Woosley,2010Smitha}), binary interactions or mergers (2012 outburst of SN~2009ip, \citealp{2013Kashi,2013Soker}), electron-capture SN explosions of 8--10~$M_{\odot}$ stars (SN~2011ht, \citealp{2013Mauerhana}).

In search of SN progenitors, late-time imaging is crucial to confirm whether the detected progenitor candidate has vanished. This is important to distinguish a genuine progenitor from other scenarios, such as a star cluster, blend of multiple objects, or chance alignment. For SNe~IIn, this is also important to confirm whether the explosion is indeed the terminal explosion of the progenitor or a non-terminal eruption that is often as extreme as a SN explosion \citep[i.e. SN impostor;][]{1995Filippenko,2000VanDyk,2002VanDyk,2011Kochanek,2012VanDyk}.

SN~2010jl is a Type~IIn SN in the galaxy UGC 5189A. 
It attracted considerable attention for its high luminosity, reaching $M_V \approx -19.9$ mag and $M_I \approx -20.5$ mag at peak \citep{2011Stoll}, which is even close to the superluminous SNe ($M_R \le -21$ mag, \citealp{2012GalYam}). 
\citet{2011Smith} identified a luminous blue object at the site of  SN~2010jl from pre-explosion images taken by the Hubble Space Telescope (HST) Wide Field Planetary Camera 2 (WFPC2).
They proposed several possible scenarios for the object, including a host stellar cluster, an extremely luminous LBV-like star in quiescence, an LBV-like star in eruption, or a combination of the above.
They claimed that any of these scenarios would indicate a very massive progenitor with $M_{\rm ini} > 30~M_\odot$ for SN~2010jl.
Later, \citet{2017Fox} re-examined the accurate position of SN~2010jl using post-explosion images taken in 2014, 2015 and 2016 by HST Wide Field Camera 3 (WFC3), which have a much higher spatial resolution than WFPC2. They found that SN~2010jl actually has an offset from the center of the previously identified object on the pre-explosion WFPC2 images. They assumed the SN progenitor was below the detection limit of the pre-explosion WFPC2 images, which indicates a faint star below 30~$M_\odot$ for the progenitor.

After \citet{2017Fox}, SN~2010jl was again observed by HST WFC3 in 2018 and 2020, almost 8 and 10 years after its explosion, respectively. This allows us to further check whether it has faded sufficiently and investigate the nature of the luminous blue object near the SN position on the pre-explosion WFPC2 images. As we shall show in this letter, the pre-explosion object is a blend of the SN progenitor and a close source in its vicinity, and the SN brightness is now fainter than pre-explosion level, confirming the candidate is the genuine progenitor that has now disappeared. We further carry out a detailed analysis of the progenitor and its environment in order to understand its properties.

Throughout this paper, we adopt the same distance modulus ($m-M = 33.45$ mag) and Milky Way extinction [$E(B-V)=0.027$ mag] as in \citet{2011Smith}, as well as a standard extinction law with $R_V = 3.1$ \citep{ebvlaw},
as suggested by \citet{2022Li}.
All magnitudes are reported in the Vega system.

\section{Data}

The site of SN~2010jl was observed before explosion by HST WFPC2 in F300W/F814W in 2001, and imaged at late time by HST WFC3 in F275W/F336W in 2014 and in F336W/F814W in 2015, 2016, 2018 and 2020. This work utilizes data from these observations except the 2014 ones due to lack of the F814W band (Table~\ref{tab:tab1}). Images calibrated by the standard pipeline were retrieved from the Mikulski Archive for Space Telescopes (MAST\footnote{\url{https://mast.stsci.edu/search/ui/\#/hst}}), and we re-drizzled the 2016 images with the \textsc{astrodrizzle}\footnote{\url{http://drizzlepac.stsci.edu/}} package with \texttt{driz$\_$cr$\_$grow = 3} because they still suffered from high cosmic-ray contamination after the standard calibration. 
We performed point-spread-function (PSF) photometry using the \textsc{dolphot} package \citep{2000Dolphin} with the drizzled images (\texttt{$*\_$drc.fits}) as reference and photometric parameters recommended in the user's guide.
In particular, we adopted \texttt{img$\_$RAper = 3}, \texttt{FitSky = 2}, and \texttt{Force1 = 1} for better a modelling of the crowded SN field.

\begin{table*}[htbp]
\centering
\caption{HST data and photometry. \label{tab:tab1}}

    \begin{tabular}{cccccccc}
    \hline
    \hline
    
     Year &  Proposal & Instrument & Filter & Exposure  &  \textit{s0}  & \textit{s1}  & \textit{s2}  \\
     & ID & & & time (s) & (mag) & (mag) & (mag) \\

     \hline
    
      2001 & 8645 & WFPC2 & F300W & 1000 & 21.23 (0.06) & - & - \\
      & & & F814W & 100  & 22.92 (0.16) & - & - \\
      2015 & 14149 & WFC3  & F336W & 710  & - & 21.12 (0.03) & 22.50 (0.15) \\
       &   &   & F814W & 780 & - &  22.29 (0.01) & 23.97 (0.11) \\
      2016 & 14668 & WFC3   & F336W & 710 & - & 22.13 (0.15) & 22.53 (0.17) \\ 
       &   &    & F814W & 780  & - & 22.63 (0.03) & 23.89 (0.17) \\ 
     2018 & 15166 & WFC3   & F336W & 780 & - & 22.81 (0.08) & 22.55 (0.04) \\ 
       &   &    & F814W & 710  & - & 23.15 (0.06) & 23.97 (0.13) \\ 
     2020 & 16179 & WFC3   & F336W & 710  & - & 23.61 (0.08) & 22.44 (0.10) \\ 
       &   &     & F814W & 780  & - & 23.50 (0.19) & 23.92 (0.04) \\  
    \hline
 \end{tabular}

     \begin{tablenotes}
\item[1] PIs: 8645: R.Windhorst; 14149, 4668, 15166, 16179:  A. Filippenko
      \end{tablenotes}

\end{table*}

\section{SN progenitor}\label{sec:prog}

\begin{figure*} 
    \centering 
    \includegraphics[width=\linewidth]{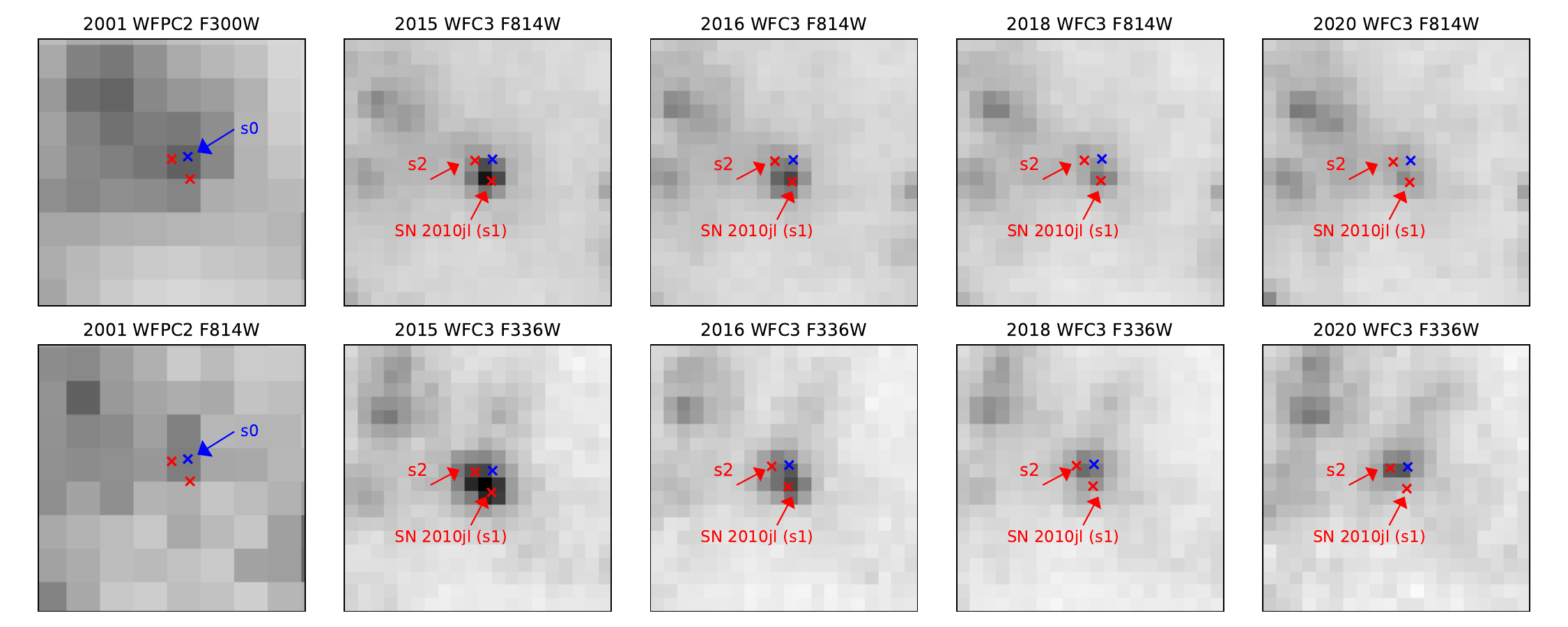}
    \caption{HST images centered at SN~2010jl. All stamps have the same dimension of 0.8" $\times$ 0.8", and are aligned with North up and East to the left. The crosses mark the positions of sources near the SN position, i.e. the pre-explosion source \textit{s0} detected on the WFPC2 images (blue), and the post-explosion sources \textit{s1} and \textit{s2} detected on the WFC3 images (red). For comparison, we have astrometrically aligned the images and overplot the positions of all three sources on each image.
    \label{fig:hst}}
\end{figure*}

\begin{figure} 
    \centering 
    \includegraphics[width=\linewidth]{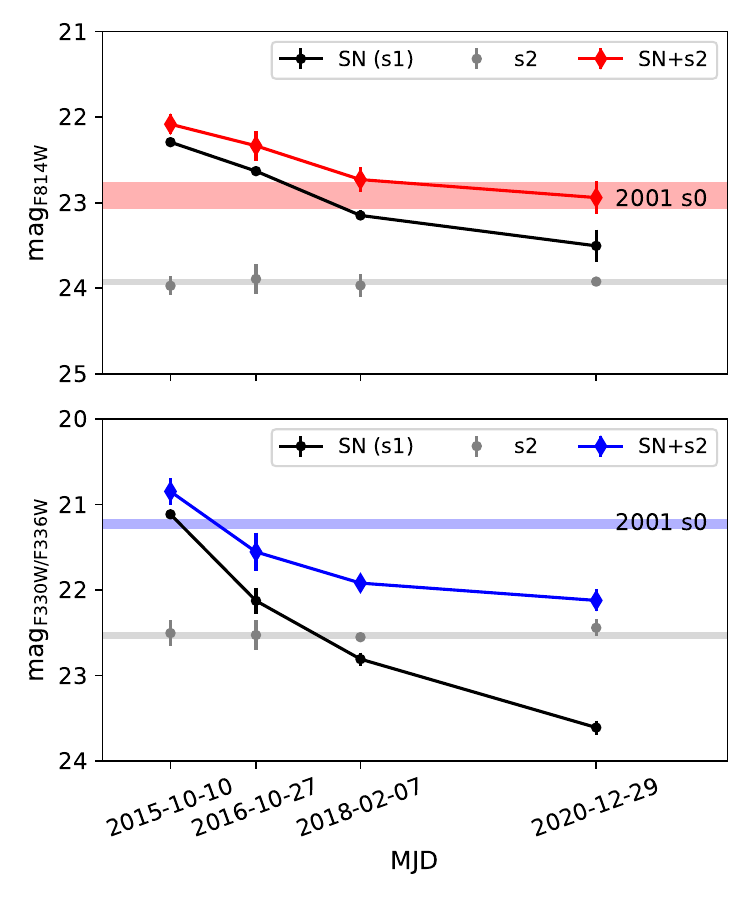}
    \caption{HST photometry of \textit{s1} (i.e. SN~2010jl; black circles) and \textit{s2} (grey circles) at late times. Their composite magnitudes are also calculated and displayed as the red or blue diamonds. The grey horizontal lines correspond to the inverse-variance weighted average magnitudes of \textit{s2} at different epochs, and the red/blue lines are the pre-explosion brightness of \textit{s0} observed in 2001. The photometric uncertainties are reflected by the error bars associated with the data points and the widths of the horizontal lines.
  \label{fig:lc}}
\end{figure}

Near the SN position (Fig.~\ref{fig:hst}), we detect a single point source on the lower-resolution pre-explosion WFPC2 images (hereafter \textit{s0}) and two spatially resolved sources on the higher-resolution late-time WFC3 images (hereafter \textit{s1} and \textit{s2}). 
The measured magnitudes of \textit{s0}, \textit{s1} and \textit{s2} are listed in Table~\ref{tab:tab1} and shown in Fig.~\ref{fig:lc}. 
While \textit{s1} continues to fade on the late-time images, \textit{s2} has almost constant magnitudes in both bands. Therefore, we identify \textit{s1} as SN~2010jl and \textit{s2} as a nearby field object. 
We note that we didn't run \textsc{dolphot} with the mode of warmstart, namely, all reported sources were detected and selected on its own.
The source identification is consistent with \citet{2017Fox}.

Image alignment was performed between the pre-explosion WFPC2 images and the late-time WFC3 images with 9 common objects, reaching a differential astrometric uncertainty of 0.028". The position of \textit{s0} agrees with that of \textit{s2} within the astrometric uncertainty (also found by \citealt{2017Fox}). The SN position (i.e. \textit{s1}) offsets from \textit{s0} by 0.08", which is larger than the 1$\sigma$ astrometric uncertainty but still smaller than the 3$\sigma$ astrometric uncertainty (0.084") and the pixel size of the WFPC2 images (0.1"). Moreover, the derived source positions have their own uncertainties. Therefore, we suggest that the positions of \textit{s0} and \textit{s1} are not inconsistent with each other and that \textit{s0} is a blend of \textit{s2} and the SN progenitor at the position of \textit{s1}.

We also carefully checked quality parameters calculated by \textsc{dolphot}.
We find \textit{s0} has \texttt{sharpness} of $-$0.257  and \texttt{roundness} of 0.154, for comparison, \textit{s2} has \texttt{sharpness} of $-$0.747, $-$0.344, $-$0.345, $-$0.293 from 2015 to 2020, and \texttt{roundness} of 1.170, 0.561, 0.496 , 0.377; \textit{s1} has \texttt{sharpness} of $-$0.262, $-$0.086, $-$0.192, $-$0.302, and \texttt{roundness} of 0.185, 0.198 , 0.419 , 0.491.
This shows \textit{s0} can't be recognized as a blend object simply based on results of WFPC2 photometry, since the offset between its two components is less than one WFPC2 pixel.

Figure~\ref{fig:lc} shows the late-time brightness evolution of SN~2010jl and \textit{s2} in comparison with the pre-explosion brightness of \textit{s0}. At these late-time epochs, SN~2010jl has faded significantly so that \textit{s2} can be clearly resolved with its magnitudes accurately measured. We find that the light of \textit{s2} is not sufficient to account for the brightness of \textit{s0}, suggesting non-negligible light contribution from the blended SN progenitor. In particular, the composite brightness of \textit{s2} and SN~2010jl in 2018 and 2020 is significantly fainter than that of \textit{s0} (although we note the filter difference between F300W and F336W). It is possible that the SN progenitor has disappeared after explosion (see Section \ref{sec:dis}).

\begin{figure*} 
    \centering 
    \includegraphics[width=0.9\linewidth]{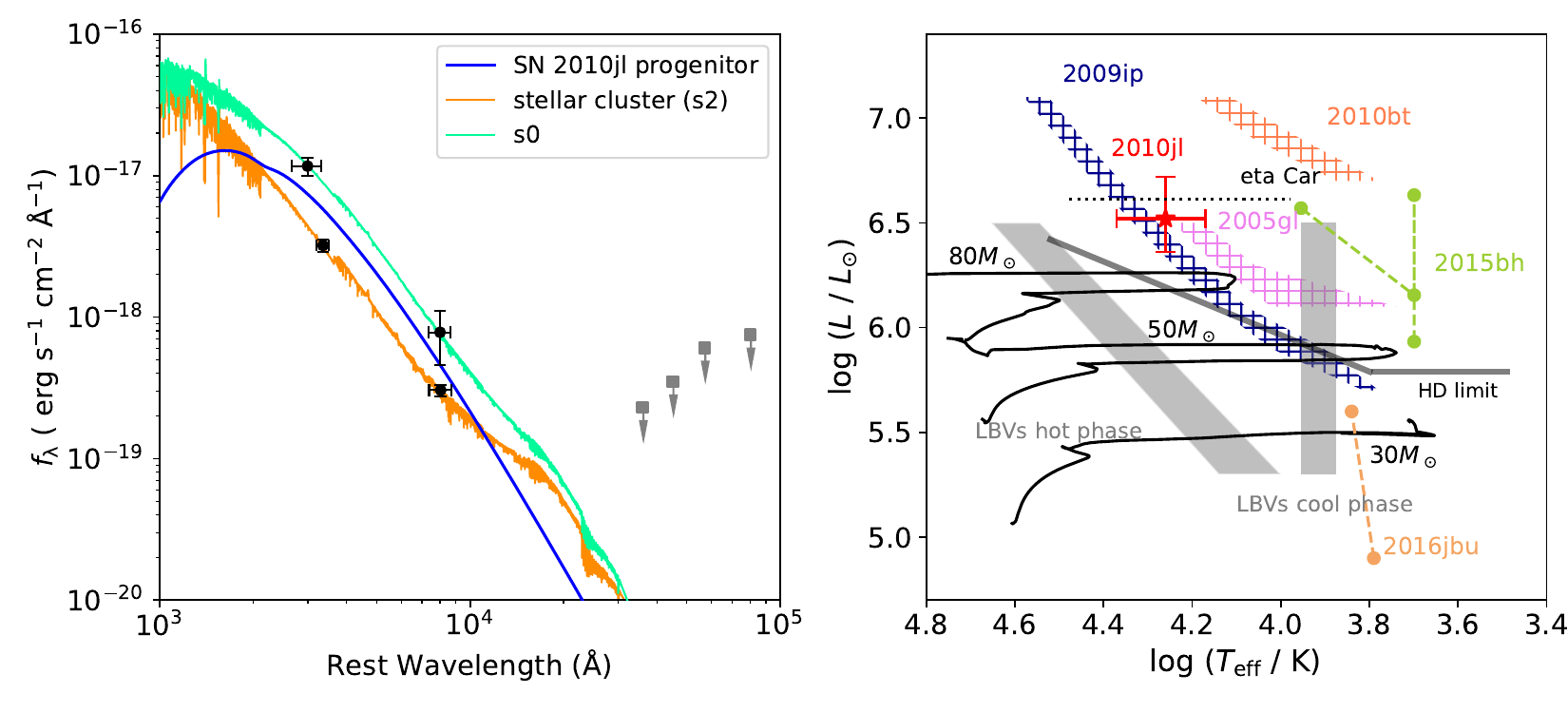}
    \caption{\textit{Left}: Photometry decomposition of the pre-explosion source \textit{s0}, which is a blend of the  progenitor of SN~2010jl (i.e. \textit{s1}; blue) and the nearby field cluster \textit{s2} (orange). The black data points are based on HST observations as derived in this work, while the grey data points are based on Spizter/IRAC observations as reported in \citet{2017Fox}.
    The progenitor is consistent with a blackbody of log($T_{\rm eff}/{\rm K}$)=4.26$^{+0.11}_{-0.09}$ and log($L/L_{\odot}$)=6.52$^{+0.20}_{-0.16}$. 
    \textit{Right}: Position of SN~2010jl's progenitor on the HR diagram (red star with error bars). The other directly detected Type~IIn SN progenitors are also displayed, including SN~2005gl \citep{2007Gal-Yam}, SN~2009ip \citep{2010Smith}, SN~2010bt \citep{2018EliasRosa}, SN~2015bh \citep{2016Elias-Rosa}, and SN~2016jbu \citep{2018Kilpatrick}. For the last two SNe (2015bh and 2016jbu), the progenitors are observed at multiple epochs and exhibit significant variability; we show their positions at all the epochs as reported in the references. For the progenitors of the first three SNe (2005gl, 2009ip and 2010bt), the effective temperatures are not well constrained by the detection (often in a single band), so we conservatively leave effective temperature as a free parameter and re-fit the reported photometry with a blackbody SED; the results are shown by the hatched regions. For comparison, We also plot single-star isochrones of $M_{\rm ini}$=30, 50, 80~$M_{\odot}$ provided by \textsc{MIST} \citep[black lines;][]{2016mist}, standard locations of the hot/cool phases for LBVs \citep[grey-shaded regions;][]{2004Smith}, the HD limit \citep[grey broken line;][]{1994Humphreys} and the position of the Galactic LBV $\eta$~Car \citep[black dotted line;][]{2012Groh,2019Mehner}.
    \label{fig:sed}}
\end{figure*}

Next we try to decompose the spectral energy distribution (SED) of \textit{s0} to isolate the SN progenitor and derive its stellar parameters. We use a blackbody to simulate the SED of the SN progenitor; as we shall show in the next section, \textit{s2} is likely a star cluster so we adopt the BPASS v2.2 population models including interacting binaries \citep{2017Eldridge} to model its SED. We assume that the SED of \textit{s0} is the sum of both components; we argue that this is a reasonable assumption since \text{s2} and the SN progenitor are so close to each other while the other sources in this field are too far away to contaminate \textit{s0} significantly.  
We only consider Galactic extinction since the host-galaxy extinction is negligible \citep{2011Smith}.
We use the \textsc{pysynphot} package \citep{pysynphot.ref} to synthesize model magnitudes and then compare them with the observed ones of \textit{s0} and \textit{s2}. The Markov Chain Monte Carlo (MCMC) technique is used \citep{2013emcee} to search for the best-fitting parameters. We find that the progenitor has an effective temperature of log($T_{\rm eff}/{\rm K}$)=4.26$^{+0.11}_{-0.09}$ and a bolometric luminosity log($L/L_{\odot}$)=6.52$^{+0.20}_{-0.16}$; \textit{s2} has an initial mass of 10$^4$~$M_\odot$ and a very young age of only 5~Myr. The best-fitting SEDs are displayed in Fig~\ref{fig:sed}. Figure~\ref{fig:sed} also shows the position of SN2010jl's progenitor on the Hertzsprung-Russell (HR) diagram. This analysis shows that the progenitor of SN~2010jl is a very blue and extremely luminous star that is even above the Humphreys-Davidson (HD) limit \citep{1994Humphreys}.

\section{SN environment} \label{sec:env}

\begin{figure*} 
    \centering 
    \includegraphics[width=0.9\linewidth]{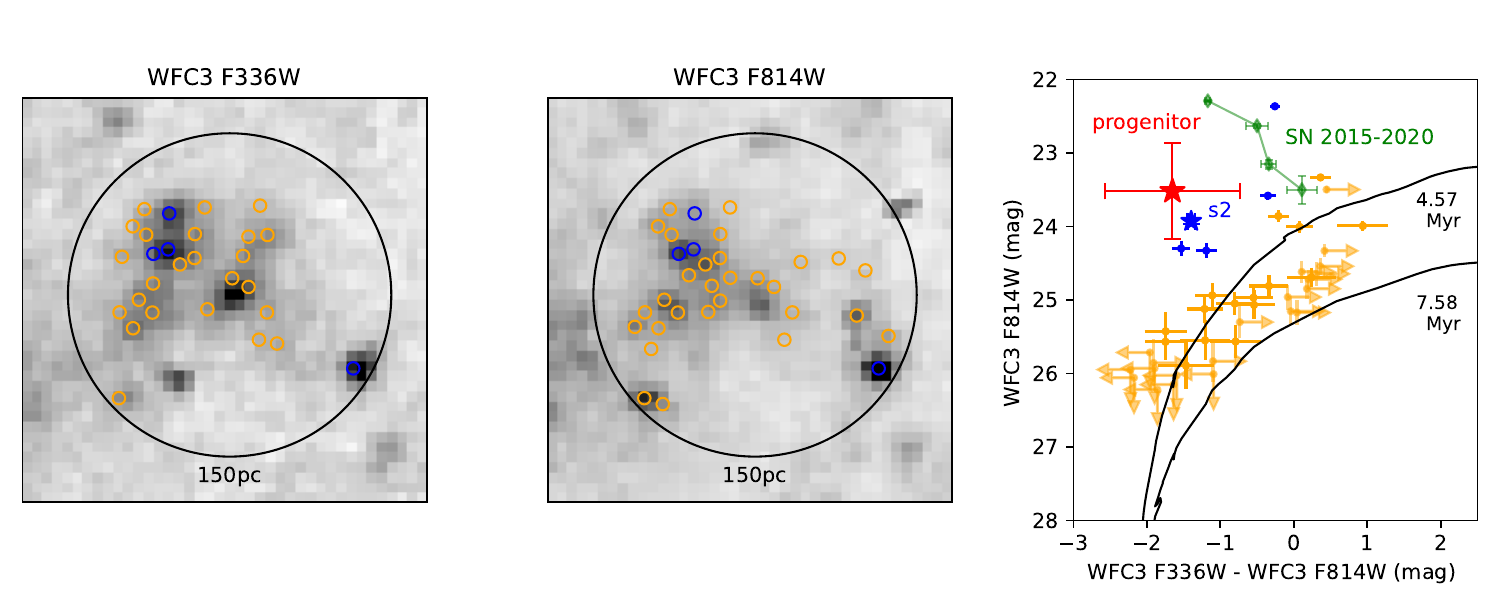}
    \caption{
    \textit{Left and middle}: WFC3 images of the SN environment with 4 different epochs combined together. Both images have the same dimension of 1.6" $\times$ 1.6" and are centered on the SN position with North up and East to the left. The small circles correspond to the detected point sources for each band within 150~pc of the SN position, and they are color-coded in the same way as in the right-hand panel.
    \textit{Right}: color-magnitude diagram of the resolved stellar population in the SN vicinity (circles). \textsc{parsec} v1.2S single-star isochrones are over-plotted, which correspond to the two best-fitting model stellar populations (see text). Sources which can be well fitted by model stellar populations are shown in orange while the over-luminous ones are displayed in blue. In addition, we also show the SN progenitor (red star; based on synthetic magnitude from our SED decomposition in Section~\ref{sec:prog}), the late-time behavior of SN~2010jl (green diamonds; which becomes fainter and redder with time), and the field cluster \textit{s2} (blue star). The error bars reflect photometric uncertainties.
  \label{fig:env}}
\end{figure*}

In this section we investigate the surrounding stellar populations in the environment of SN~2010jl. An environmental analysis can provide useful insights into the SN progenitor, since most massive stars form in groups and stars within each group have very similar ages and chemical abundances. 
We performed photometry with \textsc{dolphot} on all WFC3 images together, so that the derived photometry in each band is the inverse variance-weighted average of those on the individual frames, and then we selected point sources from the combined photometry catalog with the following criteria:

(1) \texttt{object type} = 1;

(2) \texttt{signal-to-noise ratio} $\ge$ 3;

(3) $-$0.5 $\le$ \texttt{sharpness} $\le$ 0.5;

(4) \texttt{crowding} $\le$ 2.

\noindent 42 sources are finally selected; for those detected in only one filter, we use artificial star tests at their positions to estimate the detection limits in the other filter. Figure \ref{fig:env} plots the images and the color-magnitude diagram (CMD) of the SN environment.

We apply a hierarchical Bayesian approach to fit model stellar populations to the data.
Detailed description about this method can be found in \citet{Maund2016} and \citet{Sun2021} (see also \citealp{Maund2017,Sun2020, 2022Sun,Sun2023a,Sun2023b}).
In brief, simulations of the stellar populations are based on the \textsc{parsec} v1.2S stellar isochrones \citep{parsec.ref} with solar metallicity. 
The detected sources are assumed to be single stars or non-interacting binaries from model stellar populations with different mean log-ages and same extinction, and stars within each model population has a Gaussian log-age distribution with a small dispersion of 0.05~dex. 

We find that stars in the SN environment can be fitted with two model stellar populations with mean log-ages of log($t$/yr) = 6.66 and 6.88 dex (i.e. 4.57 and 7.58~Myr, respectively). They have an extinction of $E(B-V)=0.03 \pm 0.01$ 
, which is consistent with that for the SN \citep{2011Smith}. \textsc{parsec} isochrones of the corresponding ages are overplotted on the color-magnitude diagram (Fig.~\ref{fig:env}) for comparison. This suggests that SN~2010jl is located in a very young environment with very recent star formation; therefore, its progenitor is likely to be very young and massive.


It is worth noting, however, that the SN progenitor appears much brighter than the surrounding stellar populations. It is above the isochrone of the younger population by $>$1~mag in the F814W band, and this difference cannot be accounted by (non-interacting) binarity. It is possible that the SN progenitor is not a quiescent star and may become over luminous in outburst before the final explosion.

Additionally, a few field objects, including \textit{s2}, are also much brighter than the fitted stellar isochrones; they might be star clusters, binary merger products, multiple stellar systems, and/or LBV-like stars in outburst. Assuming \textit{s2} to be a star cluster, the derived age in Section~\ref{sec:prog} is consistent with that for the younger resolved stellar population.

\section{Discussion}

Using pre-explosion WFPC2 images, \citet{2011Smith} identified the luminous blue object at the site of SN~2010jl (i.e. \textit{s0}) as its progenitor. Later, \citet{2017Fox} found the center of \textit{s0} is more consistent with that of the nearby field cluster \textit{s2} resolved on the higher-resolution WFC3 images; they assumed \textit{s0} corresponds to \textit{s2} and the SN progenitor is below the detection limit of the pre-explosion images. 
Now that new WFC3 images have been observed at later epochs, and we are able to confirm that \textit{s0} is actually a blend of \textit{s2} and the SN progenitor.

We find the SN progenitor is an LBV-like star by decomposing the SED of \textit{s0} (Section \ref{sec:prog}). In this analysis, we only corrected for the Galactic extinction since the spectra of SN~2010jl exhibit very weak  Na \textsc{i} lines from dust within the host galaxy \citep{2011Smith}.
We note, however, several works report possible evidence for pre-existing dust in the CSM \citep[e.g.][]{2011Andrews,2014Fransson}.
The SN progenitor would become even bluer and more luminous, if non-zero circumstellar extinction is taken into account into the SED analysis.

In 2018 and 2020, SN~2010jl has faded sufficiently and the composite brightness of \textit{s1} and \textit{s2} is below the pre-explosion brightness of \textit{s0} at the near-ultraviolet wavelengths (Figs.~\ref{fig:lc} and \ref{fig:sed}). 
This indicates that pre-explosion source at the SN position is not a stellar cluster or a field star but rather the SN progenitor.
It is likely that the SN progenitor has now disappeared. 
We can't rule out the possibility that SN~2010ji is not a terminal explosion and the 
pre-explosion counterpart is still alive; in this case, the star was in outburst with higher luminosity when observed by HST in 2001 and has now returned to a fainter quiescent state at late times.
It is also possible the brightness decline is (partly) due to the newly formed dust \citep{2012Smith,2013Maeda,2014Gall}.

In addition to SN~2010jl, there are another five SNe~IIn with directly detected progenitors, including SN~2005gl \citep{2007Gal-Yam}, SN~2009ip \citep{2010Smith}, SN~2010bt \citep{2018EliasRosa}, SN~2015bh \citep{2016Elias-Rosa}, and SN~2016jbu \citep{2018Kilpatrick}. Similar to SN~2010jl, all these SNe have late-time brightness below the progenitors, suggesting the progenitors have likely disappeared after the SN explosion \citep{2022Brennan, 2018EliasRosa, 2007Gal-Yam, 2022Jencson, 2022Smith}. Obvious from the HR diagram (Fig.~\ref{fig:sed}), the detected SNe~IIn progenitors exhibit significant diversity in both effective temperature and luminosity. While SN~2016jbu's progenitor has a relatively low effective temperature and luminosity and is likely to be a 18--30-$M_\odot$ yellow hypergiant or a 17-$M_\odot$ star in a binary system \citep{2018Kilpatrick,2022Brennan,2022Brennan0}, the other detected progenitors (including SN~2010jl) have extremely high luminosities above the HD limit \citep[the empirical luminosity upper limit of massive stars in hydrostatic equilibrium;][]{1994Humphreys} that are difficult to explain with standard models of stars in quiescence.

It has been noticed that the blue colors and high luminosities of the detected SNe~IIn progenitors resemble some of the known LBVs in the Milky Way and other Local Group galaxies (e.g. $\eta$~Car; \citealp{2019Mehner}), but none of the detected progenitors align well with the commonly adopted S Dor-like instability strip; \citealp{2004Smith}). The LBV giant eruptions are also similar to the SNe~IIn precursor outbursts in terms of brightness and duration \citep{2010Smith}, which can generate the dense CSM that later interact with the SN ejecta. However, LBVs were believed to be a transitional phase between the MS and the Wolf-Rayet stages instead of direct SN progenitors \citep{2003Heger}, although recent works suggest that LBVs could be very heterogeneous and have complicated origins and final fates (see \citealp{2017Smith,2019Smith} and references therein). Therefore, the relation between SNe~IIn progenitors and LBVs remains still uncertain.

The high luminosities of SN~IIn progenitors have been interpreted as them being very massive stars of several tens of and even more than a hundred solar masses \citep{2007Gal-Yam}.
It has also been suggested that the high luminosities may arise from the progenitor being in outburst rather than in quiescence. 
This is supported by the prominent variability of several SN~IIn progenitors \citep{2019Pastorello}. In this work, we find that the progenitor of SN~2010jl is significantly brighter than expected from the environmental populations, which argues against the progenitor being a quiescent star.
This is also supported by the CSM of SN~2010jl, which was estimated to have a very high mass exceeding 10~$M_\odot$ \citep{2014Ofek}.
For the typical mass-loss rate of SN~IIn progenitors ($10^{-4} \sim 10^{0}\  M_\odot$/yr, \citealp{2014Smith}), the CSM mass corresponds to a timescale of $10^{1} \sim 10^{5}$ years before explosion during which the SN progenitor was highly unstable and formed the dense CSM via mass ejections. Therefore, it is very likely that the SN progenitor, when observed in 2001 (or 9 years before explosion), was out of quiescence.

\section{Summary}\label{sec:dis}

In this paper, we perform a detailed analysis of progenitor of the Type IIn SN~2010jl based on HST images. With newer late-time WFC3 images where the SN has faded sufficiently, we are able to confirm that the pre-explosion source near the SN position on the lower-resolution WFPC2 images is actually a blend of the SN progenitor and a field star cluster. SN~2010jl is now much fainter than its progenitor; it is likely that the progenitor has now disappeared, although we cannot rule out the possibility that the progenitor is still alive and has return to quiescence or become obscured by newly formed dust.

SED analysis shows that the SN progenitor has a very high effective temperature of log($T_{\rm eff}/{\rm K}$)=4.24$^{+0.10}_{-0.08}$ and an extreme luminosity of log($L/L_{\odot}$)=6.52$^{+0.20}_{-0.16}$. The SN progenitor is located in a very young  environment, where active star formation occurred only a few Myrs ago.
However, the SN progenitor appears much more luminous than expected from the environmental stellar populations. We suggest that the SN progenitor, observed before the explosion, is most likely to be in outburst, giving rise to its extremely high luminosity. The nature of the progenitor and its detailed evolutionary history still remains to be investigated.

\section*{acknowledgments}
This work is supported by the Strategic Priority Research Program of the Chinese Academy of Sciences, Grant No. XDB0550300. ZXN acknowledges support from the NSFC through grant No. 12303039, and NCS’s research is funded by the NSFC grants No. 12303051 and No. 12261141690. JFL acknowledges support from the NSFC through grants No. 11988101 and No. 11933004 and from the New Cornerstone Science Foundation through the New Cornerstone Investigator Program and the XPLORER PRIZE. 
This research is based on observations made with the NASA/ESA Hubble Space Telescope obtained from the Space Telescope Science Institute, which is operated by the Association of Universities for Research in Astronomy, Inc., under NASA contract NAS 5–26555. These observations are associated with programs 8645, 14149, 14668, 15166, and 16179, they were obtained from the Mikulski Archive for Space Telescopes (MAST) at the Space Telescope Science Institute. The specific observations analyzed can be accessed via \dataset[DOI: V]{https://doi.org/10.17909/nrbd-qp80}.
This work made use of v2.2.1 of the Binary Population and Spectral Synthesis (BPASS) models as described in \citet{2017Eldridge} and \citet{2018Stanway}.


%
\vspace{5mm}

\bibliography{sample631}{}
\bibliographystyle{aasjournal}

\end{document}